\documentclass[trackchanges]{aastex7}

\begin{document}

\title{Heartbeat Stars Recognition Based on Recurrent Neural Networks: Method and Validation}

\author[orcid=0000-0002-8564-8193]{Min-Yu Li}
\affiliation{Yunnan Observatories, Chinese Academy of Sciences, Kunming 650216, People's Republic of China}
\email{liminyu@ynao.ac.cn} 

\author[orcid=0000-0002-5995-0794]{Sheng-Bang Qian}
\affiliation{Department of Astronomy, School of Physics and Astronomy, Key Laboratory of Astroparticle Physics of Yunnan Province, Yunnan University, Kunming 650091, People's Republic of China}
\email[show]{qiansb@ynu.edu.cn}  

\author[orcid=0000-0002-0796-7009]{Li-Ying Zhu} 
\affiliation{Yunnan Observatories, Chinese Academy of Sciences, Kunming 650216, People's Republic of China}
\affiliation{University of Chinese Academy of Sciences, No.1 Yanqihu East Road, Huairou District, Beijing 101408, People's Republic of China}
\email[show]{zhuly@ynao.ac.cn}

\author[0000-0001-9346-9876]{Wen-Ping Liao} 
\affiliation{Yunnan Observatories, Chinese Academy of Sciences, Kunming 650216, People's Republic of China}
\affiliation{University of Chinese Academy of Sciences, No.1 Yanqihu East Road, Huairou District, Beijing 101408, People's Republic of China}
\email{liaowp@ynao.ac.cn}

\author[orcid=0000-0002-8421-4561]{Lin-Feng Chang}
\affiliation{Department of Astronomy, School of Physics and Astronomy, Key Laboratory of Astroparticle Physics of Yunnan Province, Yunnan University, Kunming 650091, People's Republic of China}
\email{chang-linfeng@ynu.edu.cn}

\author{Er-Gang Zhao} 
\affiliation{Yunnan Observatories, Chinese Academy of Sciences, Kunming 650216, People's Republic of China}
\email{zergang@ynao.ac.cn}

\author[orcid=0000-0002-5038-5952]{Xiang-Dong Shi} 
\affiliation{Yunnan Observatories, Chinese Academy of Sciences, Kunming 650216, People's Republic of China}
\email{sxd@ynao.ac.cn}

\author[orcid=0000-0002-0285-6051]{Fu-Xing Li}
\affiliation{Department of Astronomy, School of Physics and Astronomy, Key Laboratory of Astroparticle Physics of Yunnan Province, Yunnan University, Kunming 650091, People's Republic of China}
\email{lfxjs66@126.com}

\author[orcid=0000-0003-0516-404X]{Qi-Bin Sun}
\affiliation{Department of Astronomy, School of Physics and Astronomy, Key Laboratory of Astroparticle Physics of Yunnan Province, Yunnan University, Kunming 650091, People's Republic of China}
\email{sunqibin@ynu.edu.cn}

\author[0009-0004-0289-2732]{Ping Li} 
\affiliation{Yunnan Observatories, Chinese Academy of Sciences, Kunming 650216, People's Republic of China}
\affiliation{University of Chinese Academy of Sciences, No.1 Yanqihu East Road, Huairou District, Beijing 101408, People's Republic of China}
\email{liping@ynao.ac.cn}

%%\collaboration{all}{The Terra Mater collaboration}

%% Use the \collaboration command to identify collaborations. This command
%% takes an optional argument that is either a number or the word "all"
%% which tells the compiler how many of the authors above the command to
%% show. For example "\collaboration[all]{(DELVE Collaboration)}" wil include
%% all the authors above this command.
%%
%% Mark off the abstract in the ``abstract'' environment. 
\begin{abstract}

Since the variety of their light curve morphologies, the vast majority of the known heartbeat stars (HBSs) have been discovered by manual inspection. Machine learning, which has already been successfully applied to the classification of variable stars based on light curves, offers another possibility for the automatic detection of HBSs. We propose a novel feature extraction approach for HBSs. First, the orbital frequencies are calculated automatically according to the Fourier spectra of the light curves. Then, the amplitudes of the first 100 harmonics are extracted. Finally, these harmonics are normalized as feature vectors of the light curve. A training data set of synthetic light curves is constructed using ELLC, and their features are fed into recurrent neural networks (RNNs) for supervised learning, with the expected output being the eccentricity of these light curves. The performance of the RNNs is evaluated using a test data set of synthetic light curves, achieving 95$\%$ accuracy. When applied to known HBSs from the OGLE, Kepler, and TESS surveys, the networks achieve an average accuracy of 86$\%$. This method successfully identifies four new HBSs within the eclipsing binary catalog of Kirk et al. The use of orbital harmonics as features for HBSs proves to be a practical approach that significantly reduces the computational cost of neural networks. RNNs show excellent performance in recognizing this type of time series data. This method not only allows efficient identification of HBSs but can also be extended to recognize other types of periodic variable stars.

\end{abstract}

%% Keywords should appear after the \end{abstract} command. 
%% The AAS Journals now uses Unified Astronomy Thesaurus (UAT) concepts:
%% https://astrothesaurus.org
%% You will be asked to selected these concepts during the submission process
%% but this old "keyword" functionality is maintained in case authors want
%% to include these concepts in their preprints.
%%
%% You can use the \uat command to link your UAT concepts back its source.

%% From the front matter, we move on to the body of the paper.
%% Sections are demarcated by \section and \subsection, respectively.
%% Observe the use of the LaTeX \label
%% command after the \subsection to give a symbolic KEY to the
%% subsection for cross-referencing in a \ref command.
%% You can use LaTeX's \ref and \label commands to keep track of
%% cross-references to sections, equations, tables, and figures.
%% That way, if you change the order of any elements, LaTeX will
%% automatically renumber them.

\section{Introduction}\label{sect:introduction}

Heartbeat stars (HBSs) are a subclass of detached ellipsoidal variables with eccentric orbits. Their name derives from the presence of the ``heartbeat'' feature, similar to an electrocardiogram, in their light curves \citep{2012ApJ...753...86T}. The components are distorted by the time-varying tidal potential, and their response is usually divided into two parts: the equilibrium tide and the dynamical tide. The equilibrium tide is responsible for the heartbeat signature near periastron, while the dynamical tide induces the tidally excited oscillations (TEOs; \citet{1975A&A....41..329Z,1995ApJ...449..294K,2017MNRAS.472.1538F,2021FrASS...8...67G,2022A&A...659A..47K}). HBSs are rare and interesting objects that are ideal laboratories for studying celestial activities such as the evolution of eccentric orbits \citep{2012ApJ...753...86T, 2016ApJ...829...34S, 2023ApJS..266...28L}, stellar evolution \citep{2014AA...564A..36B, 2021MNRAS.506.4083J, 2025arXiv250317133M}, theoretical work on TEOs \citep{2017MNRAS.472.1538F, 2021FrASS...8...67G, 2023A&A...671A..22K}, pulsation phases and mode of TEOs \citep{2014MNRAS.440.3036O, 2020ApJ...888...95G, 2024ApJ...974..278L, 2024MNRAS.530..586L}, resonance locking \citep{2017MNRAS.472L..25F, 2018MNRAS.473.5165H, 2020ApJ...903..122C}, apsidal motion \citep{2016MNRAS.463.1199H, 2021ApJ...922...37O}, massive HBSs and exceptional objects \citep{2022A&A...659A..47K, 2024A&A...686A.199K, 2023NatAs...7.1218M, 2024A&A...685A.145K}, and hybrid pulsation systems \citep{2013MNRAS.434..925H, 2019ApJ...885...46G,2025PASJ...77..118L}, etc.

It was only with the release of photometric data from long-term baseline and high-precision surveys that HBSs were discovered in large numbers. \citet{2016AJ....151...68K} reported the first large catalog of Kepler HBSs with 173 systems, consisting of some objects reported in previous literature and those they discovered by manual classification. \citet{2021A&A...647A..12K} found 20 HBSs from the Transiting Exoplanet Survey Satellite (TESS) data by visual inspection. \citet{2022ApJS..259...16W} reported the largest catalog of HBSs to date, with 991 objects discovered from the Optical Gravitational Lensing Experiment (OGLE) database through the visual inspection of experienced researchers. \citet{2024MNRAS.534..281L} discovered 23 TESS HBSs by visual search of light curves. Recently, \citet{2025ApJS..276...17S} identified 180 TESS HBSs constructed in two parts. One part, with a larger number, was selected manually from the light curves identified by a neural network trained to search for eclipsing binaries. The other 33 HBSs were also manually inspected from other candidates identified by another neural network trained to search for HBSs. In summary, the vast majority of known HBSs have been discovered by manual inspection, even with some machine learning technology. Therefore, a powerful automatic detection algorithm is needed to reduce the need for manual inspection.

Machine learning has demonstrated remarkable success in classifying variable stars using light curve data\citep{2018NatAs...2..151N}, establishing it as a robust approach for the automated detection of HBSs. Furthermore, feature extraction plays a pivotal role in the machine learning-based classification of light curves \citep{2020ApJS..250...30J}, with various methodologies being employed in practical implementations. One intuitive approach is to mathematically calculate a set of well-defined features from the light curve. A number of papers used this approach, and the difference between them is that they defined different features (e.g., \citet{2018AJ....156....7H, 2019MNRAS.483.5534S, 2021AJ....162..209A, 2021AJ....161..141S, 2023AJ....166..189E, 2024MNRAS.528.6997G, 2025ApJS..276...57G}). 

Some other works developed their own algorithms to extract light curve features. \citet{2016MNRAS.456.2260A} used SOM, an unsupervised machine learning algorithm, to derive the features. \citet{2019ApJ...877L..14T} used a recurrent neural network auto-encoder for unsupervised feature extraction. \citet{2020MNRAS.491.3805J} exploited two novel time domain feature transforms, SSMM and DF, for feature extraction. \citet{2022A&A...666A.122M} used color indices as features to classify massive stars. \citet{2022MNRAS.514.2793B} extracted the features using HCTSA, an open-source software package.

With the success of convolutional neural networks (CNNs) in image identification, some works also used CNNs to recognize the corresponding images of light curves, such as images of phase-folded light curves and/or frequency spectra (e.g., \citet{2018MNRAS.476.3233H, 2018ApJ...859...64H, 2020A&A...633A..53O, 2024ApJS..274...29C,2025A&A...695A..81U}), the flux heatmaps (two-dimensional images created from flux values) \citep{2021AJ....162...67Q}, or the 2D histograms of the light curves \citep{2024A&A...691A.106M}. Moreover, Some works fed raw or transformed light curves into the neural network for learning without feature extraction (e.g., \citet{2019MNRAS.482.5078A, 2020MNRAS.493.2981B, 2021MNRAS.505..515Z, 2021A&A...652A.107V}). In these cases, the feature extraction is implemented implicitly within the neural networks. 

However, most automatic classification methods for variable stars only effectively classify stars into well-known classes such as EA, EB, RRab, etc., while HBSs are a subclass of ellipsoidal variables. The variety of their light curve morphologies makes it challenging to identify them automatically. Going back to \citet{2016AJ....151...68K}, it can be noted that they used the t-Distributed Stochastic Neighbor Embedding (t-SNE) technique to perform a visualization of their objects. It was found that good distributions were obtained for all types of stars except HBSs, while most of these HBSs are located in a specific region composed of other noisy or unique light curves (see their section 7.2 and Figure 4 panel (b) for details). This also implies that HBSs are difficult to classify automatically. On the other hand, the work of the aforementioned \citet{2025ApJS..276...17S} is a useful experiment. They trained a 12-layer 1D CNN to detect HBSs and manually identified 33 HBSs from the top 100 results of the CNN detection. This indicates that neural networks can also be used to recognize HBSs.

The present study was inspired by the unique features of the harmonics in the Fourier spectra of the light curves found when analyzing TEOs in HBSs in our previous work\citep{2024ApJ...962...44L}. As periodic variables, the harmonics of HBSs can be used to sufficiently characterize their light curves in the frequency domain, while other frequencies can be neglected. This can significantly reduce the amount of data for the features and then effectively reduce the learning cost of the neural network.

The paper is organized as follows. Sect.~\ref{sect:methodology} details the neural network methodology and training procedures. Sect.~\ref{sect:results} assesses the networks' generalization capability and evaluates their performance on actual light curve data. The results are analyzed in Sect.~\ref{sect:discussion}, followed by a summary and conclusions in Sect.~\ref{sect:sc}.

\section{Methodology}\label{sect:methodology}

\subsection{Samples generation}\label{sect:samples}
To obtain high-resolution Fourier spectra from synthetic light curves, the data must contain a sufficient number of points and span multiple orbital periods. To this end, ELLC\citep{2016A&A...591A.111M} is ideal for generating synthetic light curves of HBSs because it allows precise control over temporal sampling and orbital phase coverage.

To generate light curves with different morphological characteristics, we systematically varied seven main parameters: orbital eccentricity ($ecc$), orbital inclination ($incl$), argument of periastron ($\omega$), mass ratio ($q$), surface brightness ratio ($sratio$), and radius in units of the semi-major axis of the primary and secondary ($r_1$ and $r_2$). For other additional parameters, including the geometric albedos, limb and darkening coefficients, etc., we adopt their default values.

\begin{figure*}
	\centering
	\includegraphics[width=\hsize]{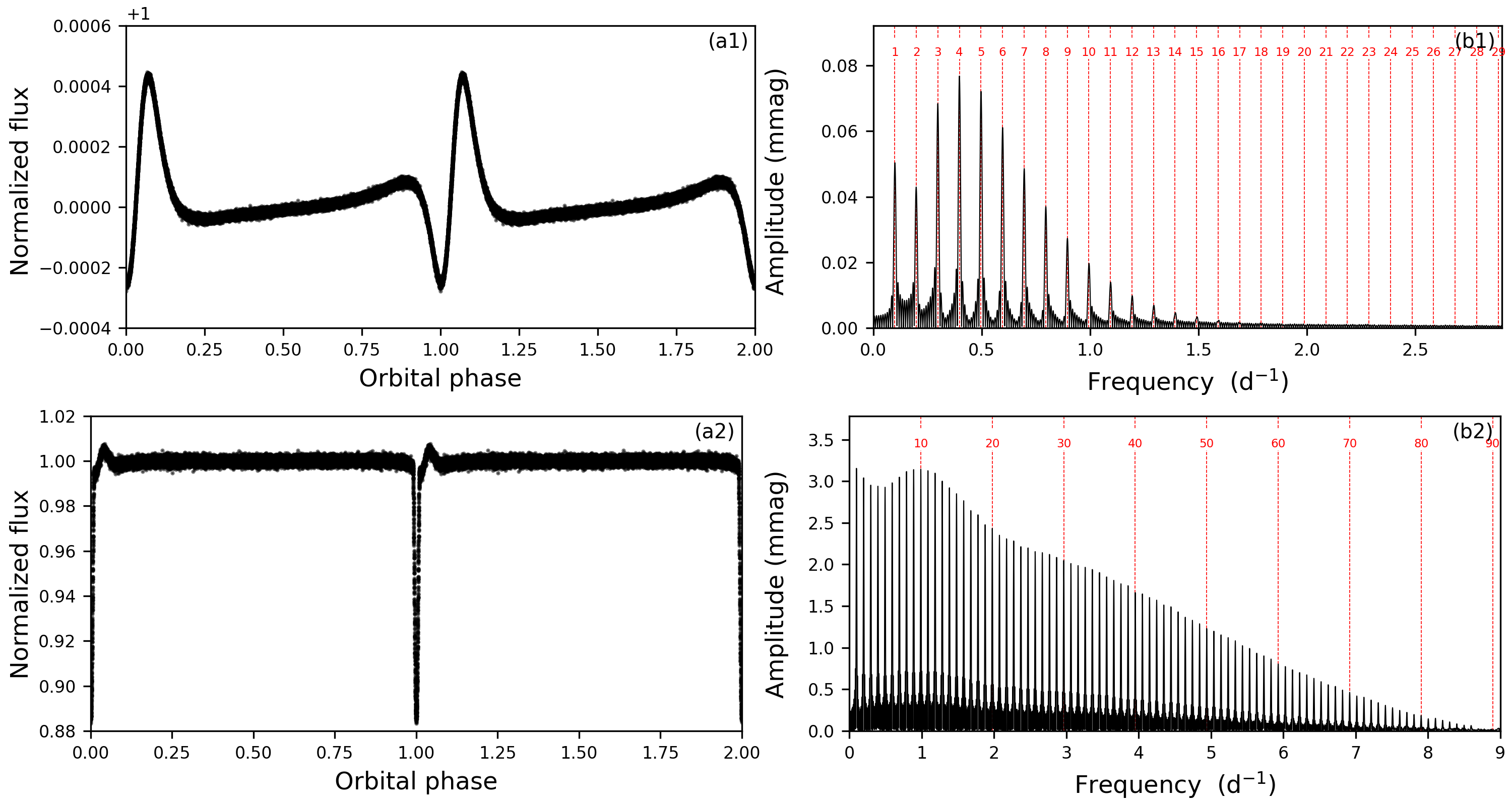}
	\caption{Two examples of the synthetic light curves and their Fourier spectra. Panels (a1) and (a2) show the non-eclipsing and eclipsing HBSs, respectively. Panels (b1) and (b2) show the corresponding Fourier spectra. The red vertical dashed lines represent the orbital harmonics.}
	\label{fig:LCs}
\end{figure*}

The seven main parameters have the following value ranges: $ecc$ is 0 to 0.9, $incl$ is 20$^{\circ}$ to 90$^{\circ}$, $\omega$ is 0$^{\circ}$ to 360$^{\circ}$, $q$ is 0.1 to 3.0, $sratio$ is 0.3 to 2.0, and $r_1$ and $r_2$ are both 0.01 to 0.1. In principle, such a parameter range setting will cover the vast majority of the known HBSs. We randomly select a value for each parameter within the corresponding range to generate a synthetic light curve. Each light curve contains about 30,000 data points and covers 10 orbital periods. We also superimpose a random flux deviation as Gaussian noise on each data point. Note that the orbital period is set to a random value, as this parameter is non-critical. Sect.~\ref{sect:feature} will show that this value will be ignored after extracting features. We generate 50,000 light curve samples using the strategy above. Panels (a1) and (a2) in Fig.\,\ref{fig:LCs} show examples of synthetic light curves for non-eclipsing and eclipsing HBSs, respectively.

The aforementioned samples consist primarily of detached binary systems with eccentricities greater than zero. However, real light curves often contain EW-, EB-, and EA-type eclipsing binaries with zero eccentricity. To prevent the neural network from misidentifying these eclipsing binaries as HBSs, an additional 2,000 semi-detached and close binaries are generated with the following parameter ranges: $ecc$ and $\omega$ are fixed at 0, while $incl$ is 60$^{\circ}$ to 90$^{\circ}$, $q$ is 0.1 to 1, $sratio$ is 0.2 to 0.5, and $r_1$ and $r_2$ are both 0.1 to 0.5. In total, we have obtained 52,000 samples for neural network training.

%% The "ht!" tells LaTeX to put the figure "here" first, at the "top" next
%% and to override the normal way of calculating a float position.
%% The asterisk after "figure" tells the compiler to span multiple columns
%% if a two column style is selected.

\subsection{Features extraction}\label{sect:feature}
The data points of a light curve and its Fourier spectrum are mathematically equivalent. For HBSs, the periodicity features are determined solely by the harmonics in the Fourier spectra. Therefore, the harmonics can fully characterize the light curve features. This approach significantly reduces the dimensionality of the feature space. For light curves with arbitrary orbital periods, the extracted features become an ordered set of harmonic amplitudes, with the orbital period parameter intentionally excluded. Our approach will utilize the first 100 harmonics as features of the light curve, rather than employing all harmonics. Sect.\,\ref{sect:training} will demonstrate that this is an effective approximation.

We use the FNPEAKS \footnote{\url{http://helas.astro.uni.wroc.pl/deliverables.php?active=fnpeaks}} code to perform the Fourier transform of the synthetic light curve data. This software significantly reduces computation time, accurately extracts individual peak frequencies, and derives their signal-to-noise ratios ($S/N$), where noise is defined as the average of all peak frequencies.

The frequency $f$ with $S/N \ge 4.0$ can be considered a harmonic if it satisfies the following equation:
\begin{equation}\label{equation:a}
	|n-f/f_{\rm orb}| < 0.05,
\end{equation}
where $n$ is the harmonic number, $f_{\rm orb}=1/P$ is the orbital frequency, and $P$ is the orbital period. To more comprehensively extract harmonic signals, we appropriately relaxed the decision threshold (compared with Eq. (1) in \citet{2024ApJ...962...44L}).

Each synthetic light curve is transformed into a set of eigenvectors:
\begin{equation}\label{equation:b}
	A = (a_1, a_2,..., a_n),
\end{equation}
where $a_i$ denotes the amplitude of harmonic $i$. Panels (b1) and (b2) in Fig.\,\ref{fig:LCs} show the harmonics of the corresponding HBSs as examples.

\subsection{Neural networks training}\label{sect:training}

\begin{figure}
	\centering
	\includegraphics[width=0.4\hsize]{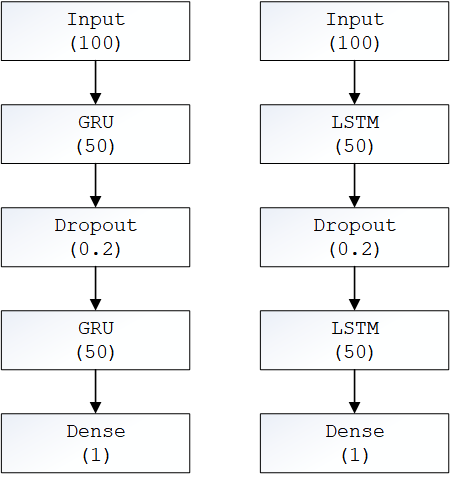}
	\caption{The architecture of the two neural networks. Each network consists of two recurrent layers (either GRU or LSTM) of 50 units each, interleaved with a dropout layer (rate=0.2). The input layer processes 100-dimensional features, and a one-unit dense layer generates the output.}
	\label{fig:nn}
\end{figure}

\begin{figure*}
	\centering
	\includegraphics[width=\hsize]{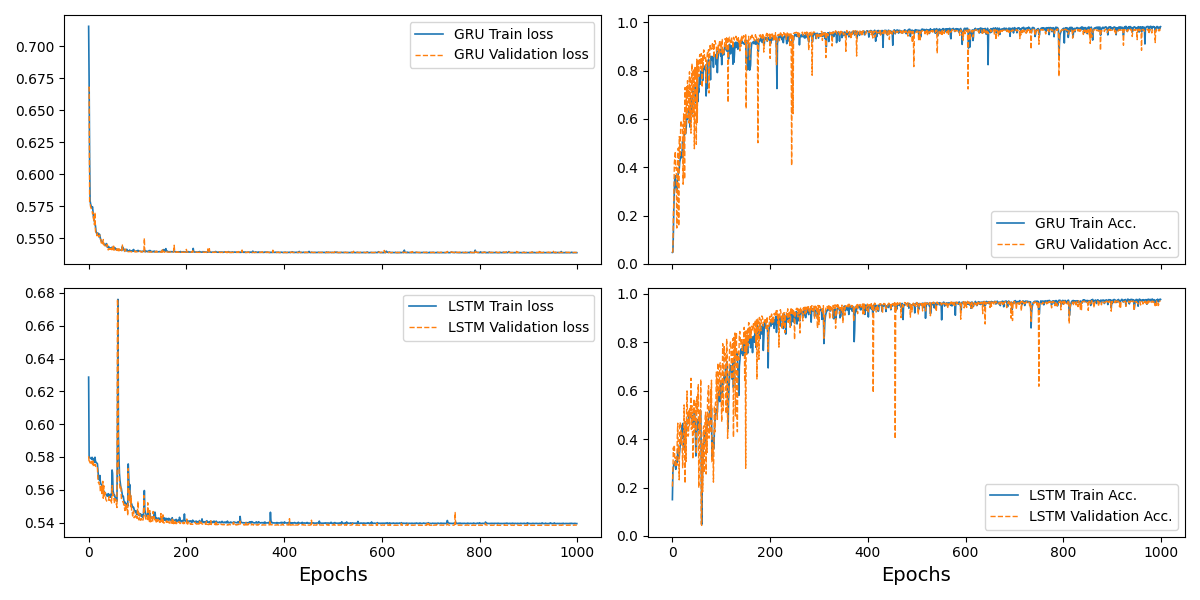}
	\caption{Training process for the GRU and LSTM networks. The left and right two panels show the loss and accuracy curves, respectively. Blue solid and orange dashed lines correspond to the training and validation sets, respectively.}
	\label{fig:loss_acc}
\end{figure*}

Recurrent neural networks (RNNs)\footnote{\url{http://karpathy.github.io/2015/05/21/rnn-effectiveness/}} are specifically designed for sequential information processing, such as time series data, and have demonstrated excellent performance in such applications. As described in Sect.~\ref{sect:feature}, the synthetic light curve of an HBS is characterized by an ordered set of harmonic amplitudes. These feature data are, therefore, particularly suitable for processing with RNNs.

Since the feature dimensions have been reduced, a highly deep neural network architecture becomes unnecessary. As shown in Fig.\,\ref{fig:nn},  we build two neural networks with two Gated Recurrent Unit (GRU) \citep{2014arXiv1406.1078C} or Long Short-Term Memory (LSTM) \citep{Hochreiter1997LSTM} layers of 50 units each and a dropout layer between them with a dropout rate of 0.2 to prevent overfitting. The output is generated by a one-unit dense layer with a ReLU \citep{2015arXiv150201852H} activation function.

Orbital eccentricity is a critical parameter for HBSs. The ``heartbeat'' signals in their light curves originate from equilibrium tides induced by time-varying tidal forces in eccentric orbits. Therefore, the one output unit is appropriate to characterize eccentricity. Moreover, since the output is a float value, a deviation threshold of 0.02 from the expected values is used to determine equivalence during training.

The input layer contains 100 units. We note that non-eclipsing HBSs typically have fewer than 30 effective harmonics (see panel (b1) in Fig.\,\ref{fig:LCs}), while most eclipsing systems exhibit fewer than 100 harmonics (see panel (b2) in Fig.\,\ref{fig:LCs}), with only a minority reaching 150. Through extensive testing, we have found that 100 input units provide optimal performance, as larger numbers do not yield improvements. For samples with over 100 harmonics, we suggest that the first 100 harmonics contain sufficient feature information. Furthermore, the feature values must be normalized using the following equation before being input into the neural network:
\begin{equation}\label{equation:normalize}
	a_i = \frac{a_i}{\|A\|}, \quad \text{where} \quad \|A\| = \sqrt{\sum_{i=1}^n a_i^2} .
\end{equation}

The samples generated in Sect.\,\ref{sect:samples} are randomly divided into training, validation, and test sets in a ratio of 8:1:1. Fig.\,\ref{fig:loss_acc} illustrates the training process of the two networks. After about 400 epochs, both networks show lower validation accuracy than training accuracy, suggesting the beginning of overfitting. Nevertheless, they maintain over 95$\%$ testing accuracy. To ensure optimal performance, we save the network weights whenever the validation accuracy exceeds the training accuracy during the 1000-epoch training. The generalization ability of these networks will be evaluated using real light curve data in Sect.\,\ref{sect:test}.

\subsection{Orbital frequency calculation}\label{sect:derive_forb}
To apply the trained neural network to real light curve data with unknown orbital periods, we have developed a program that efficiently derives the orbital harmonic using the peak frequencies ($S/N \ge 4.0$) extracted by FNPEAKS. Our algorithm incorporates two strategies. First, we note that the most prominent frequency is usually an integer multiple of the orbital orbital frequency. Thus, we sequentially assume it to be the first, second, third, and so on harmonic order. Then, we search for  harmonics among the peak frequencies based on each assumed harmonic order. The value with the most matches across all results is identified as the actual orbital frequency. This approach generally covers most cases.

However, for systems with self-excited pulsations, such as KIC 4949187 (see Fig. A2 in \citet{2024ApJ...962...44L}) and KIC 7914906 (see Fig. 3 in \citet{2025PASJ...77..118L}), the dominant frequency is not an integer multiple of the orbital frequency. In these cases, we implement a second strategy. First, we compute the intervals between all adjacent peak frequencies. Then, we identify the interval value with the highest occurrence rate (taking frequency resolution into account) and designate it as the orbital frequency.

\section{Results}\label{sect:results}

\subsection{Test on Real Light Curves}\label{sect:test}
We test the neural networks saved in Sect.\,\ref{sect:training} on four groups of real light curve data sets to find the best-performing networks. Using the approach in Sect.\,\ref{sect:derive_forb}, we derive the orbital frequencies of the Kepler, TESS, and eccentric binary samples and achieve 90$\%$ accuracy, with fractional errors of less than 0.05 from the correct values reported in the literatures. Due to the limited sampling accuracy, however, this approach is not applicable to the OGLE samples. Therefore, we adopt the orbital frequency values of the OGLE samples from the literature. Then, we use the approach in Sect.\,\ref{sect:feature} to extract the sample features. 

\begin{figure*}
	\centering
	\includegraphics[width=\hsize]{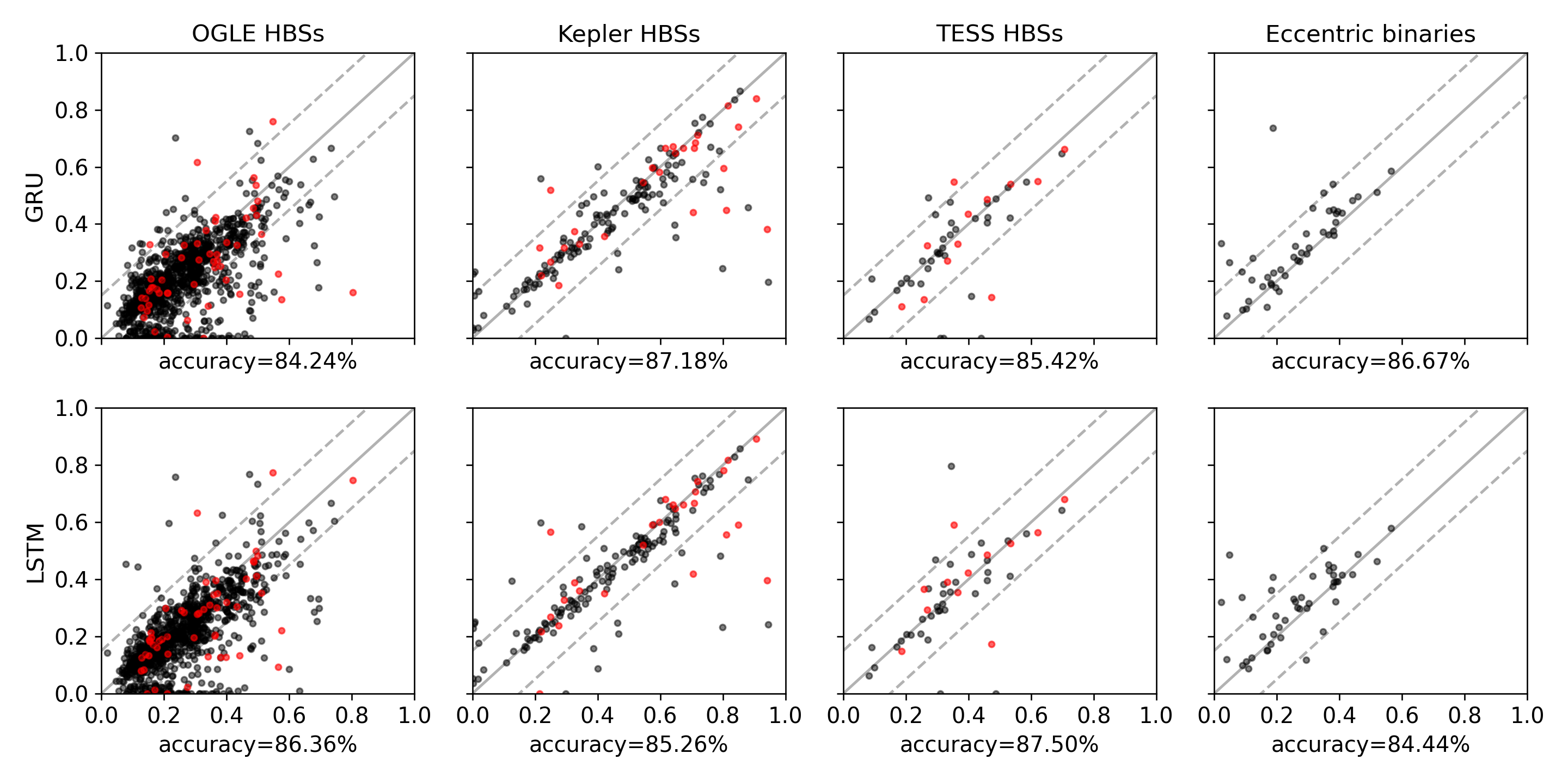}
	\caption{The first and second rows show the test results of the GRU and LSTM networks, respectively, on real data sets. The column headings denote the names of the four test data sets. In each panel, the x-axis represents the eccentricity from the reference papers;  the y-axis represents the eccentricity predicted by the RNN; the two gray dashed lines indicate deviations of $\pm$0.15. The red circles represent the HBSs with TEOs.}
	\label{fig:results}
\end{figure*}

The first and second rows of Fig.\,\ref{fig:results} show the optimal performance of the GRU and LSTM networks, respectively. The two networks achieve 86$\%$ accuracy prediction consistency (with a deviation of less than 0.15) on the four test data sets.

The sample of OGLE HBSs includes 991 systems from \citet{2022ApJ...928..135W, 2022ApJS..259...16W}. For Kepler HBSs, we adopt the objects of \citet{2023ApJS..266...28L} after excluding five systems with weak heartbeat signals (whose Fourier spectra show unreliable harmonic features), while including eight eclipsing HBSs: KICs 3230227 \citep{2017ApJ...834...59G}, 3766353 \citep{2021MNRAS.508.3967O}, 4142768 \citep{2019ApJ...885...46G}, 5790807 and 6117415 \citep{2020ApJ...903..122C}, 7914906 \citep{2025PASJ...77..118L}, 9540226 \citep{2018MNRAS.476.3729B}, and 10614012 \citep{2014AA...564A..36B}. The TESS HBS samples include 48 systems reported in \citet{2021A&A...647A..12K, 2024ApJ...974..278L, 2024MNRAS.534..281L}.

In addition, our neural networks also demonstrate effective recognition of eccentric binaries. These systems have two eclipses, with the secondary eclipse occurring at phases other than 0.5. We compile the eccentric binary test set of 45 systems from \citet{2014A&A...563A..59M, 2014MNRAS.443.3068B, 2015PASA...32...23K, 2015BlgAJ..23...75K, 2016NewA...48...30K, 2016Ap&SS.361..132K, 2016AJ....152..189K, 2016ApJ...818..108R, 2016MNRAS.461.2896H, 2016ApJ...832..121G}.

\subsection{Four new HBSs in the Kirk et al. EB catalog}\label{sect:new_hbs}
\citet{2016AJ....151...68K} have reported a catalog of 2,878 Kepler eclipsing binary (EB) systems, including eclipsing and ellipsoidal binaries, HBSs, and eccentric binaries. We collect 2,623 systems from this catalog for further testing, excluding the known HBS systems and those from which no reliable harmonic features could be extracted. We also derive the orbital frequencies of the samples following Sect.\,\ref{sect:derive_forb} and achieve an accuracy of approximately 78$\%$, with fractional errors of less than 0.05 from the correct values reported by \citet{2016AJ....151...68K}. The decrease in accuracy compared to previous samples is primarily due to twofold deviations in the orbital frequencies of some EW-type eclipsing binaries and systems with low $S/N$ spectra. We then extract the sample features following Sect.\,\ref{sect:feature}.

These sample features are identified simultaneously using the two networks in Fig.\,\ref{fig:results}. For a given sample, if both networks predict its eccentricity to be greater than 0.1, it is classified as either an HBS or an eccentric binary candidate. Conversely, if both networks predict an eccentricity less than 0.1, it is identified as a circular-orbit eclipsing binary or a contact binary. Otherwise, the average of the two network predictions is used as the final eccentricity value for subsequent classification (this situation occurs in about 8$\%$ of cases).

We then identify nearly 1,200 objects with an eccentricity greater than 0.1. Through visual inspection of their phase-folded light curves, we first exclude approximately 280 objects with incorrectly determined orbital periods. To further classify the remaining approximately 900 objects as HBSs or eccentric binaries, we implement a simple strategy: scanning the phase-folded light curves to count the number of eclipses. Systems exhibiting no eclipses or only one eclipse are classified as HBSs, while those showing more than one eclipse are identified as eccentric binaries. Note that this classification strategy is not entirely rigorous and will require subsequent visual inspection.

\begin{figure*}
	\centering
	\includegraphics[width=\hsize]{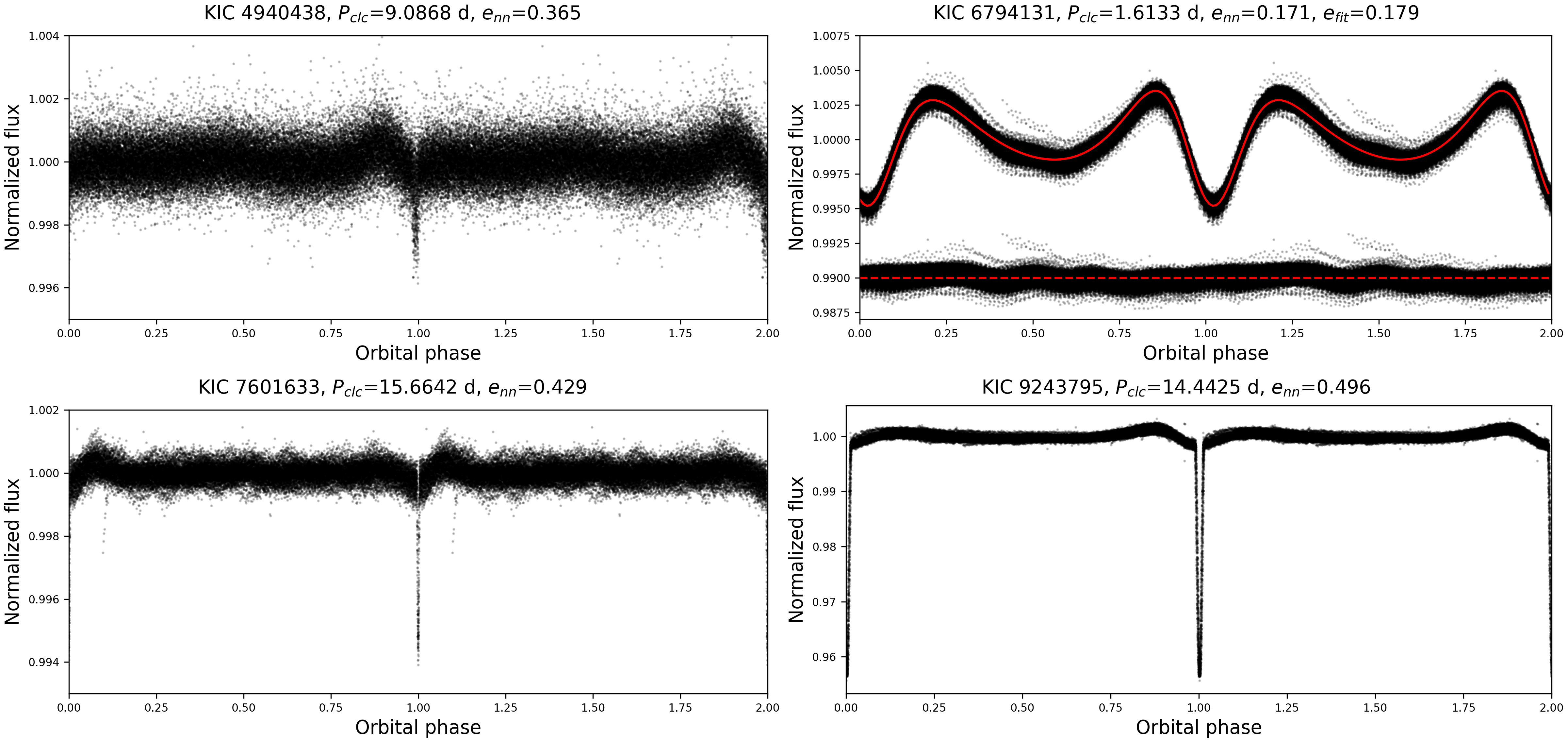}
	\caption{Four new HBSs. The orbital period ($P_{clc}$) is derived using the approach in Sect.\,\ref{sect:derive_forb}. The eccentricity ($e_{nn}$) is predicted by our RNNs. For KIC 6794131, the solid red line shows the K95$^+$ model fitted to the light curves, and the residuals are also shown. The eccentricity ($e_{fit}$) is obtained according to the K95$^+$ model. Note: The plots show phases 0$-$2 to make the heartbeat signal clear, and phases 1$-$2 are an exact copy of phases 0$-$1.}
	\label{fig:new_hbs}
\end{figure*}

Finally, among systems with eccentricities greater than 0.1, we identify about 40 systems with no eclipses, one of which is confirmed by visual inspection as a newly discovered HBS. In addition, we find nearly 420 binary systems with only one eclipse, most of which have no obvious heartbeat signals, but three of which are newly identified as HBSs. Fig.\,\ref{fig:new_hbs} shows the four new HBSs found in the \citet{2016AJ....151...68K} EB catalog, including KICs 4940438, 6794131, 7601633, and 9243795. Their orbital periods are derived using the approach in Sect.\,\ref{sect:derive_forb}, and the eccentricities are predicted by our RNNs. We further apply the method of \citet{2023ApJS..266...28L} to fit the K95$^+$ model \citep{1995ApJ...449..294K, 2022ApJ...928..135W} to KIC 6794131, as this system does not exhibit eclipse and is suitable for rapid model fitting. The derived eccentricity ($e_{fit}$=0.179) is close to the predicted value ($e_{nn}$=0.171).

\section{Discussion}\label{sect:discussion}
TEOs are induced by the dynamical tide in eccentric orbits and can be a probe to the stellar interiors. They can be observed in some of HBSs. An interesting finding is that we do not specifically construct training samples with TEOs in Sect.\,\ref{sect:samples}, but the real light curves in Fig.\,\ref{fig:results} already contain all known HBSs with TEOs as red circles. The results show that the method also has an effective prediction capability, even for HBSs with TEOs. The presence of TEOs can amplify the amplitudes of corresponding harmonics. However, we normalize these amplitudes using Eq. (\ref{equation:normalize}) to align with the input requirements of the neural network. This normalization also reduces the amplitude variations. We propose that the neural network primarily recognizes the relative magnitude of harmonics and is insensitive to abrupt changes in individual harmonics. Consequently, the neural networks can effectively generalize the TEO-related information embedded in the harmonics.

Using harmonics as features for light curve samples has at least two advantages. First, it effectively reduces the number of features, thereby decreasing the computational cost of the neural network. Second, this method inherently disregards the orbital period information of celestial objects, allowing the identification of HBSs with arbitrary orbital periods. Through comparative testing, we employ the first 100 harmonics as features of the light curves, a method that proves to be effective. In addition, the approach can be extended to other types of periodic variable stars since the orbital harmonics contain sufficient periodic features.

Eccentric binaries typically exhibit two eclipses, with the secondary eclipse significantly deviating from a phase of 0.5. Like HBSs, they have eccentric orbits. Unlike HBSs, however, they lack prominent heartbeat signals and can only be detected when their orbital inclination is large enough to produce two eclipses. Our neural network can properly predict the eccentricity of both types of systems, but further visual inspection is required for identification.  Some systems, such as KIC 4544587\citep{2013MNRAS.434..925H, 2021ApJ...922...37O}, exhibit both two eclipses and heartbeat signals. In such cases, we suggest classifying them as either type.

Sect.\,\ref{sect:new_hbs} shows that our method still requires some visual inspection effort to identify HBSs since systems containing only one eclipse do indeed exhibit high eccentricity. However, the visual inspection effort is significantly reduced compared to previous approaches. For example, we have identified nearly 30 HBSs by visual inspection of 160,000 TESS objects \citep{2024ApJ...974..278L, 2024MNRAS.534..281L}.

\section{Summary and conclusions}\label{sect:sc}
Feature extraction is a critical aspect of the application of machine learning. We propose a novel feature extraction approach for HBSs. First, the light curve is transformed into a Fourier spectrum and the orbital frequency is calculated automatically. Then, the amplitudes of the first 100 harmonics are extracted. Finally, these harmonics are normalized as feature vectors of the light curve for neural network identification. The features of the synthetic light curves are fed into a neural network for supervised learning, with the desired output being the eccentricity of the light curves. Our main results can be summarized as follows.

\begin{enumerate}
	\item We have developed a program for calculating orbital frequencies from light curves. This program is particularly effective for analyzing long-term, high-precision photometric data from missions such as Kepler and TESS.
	\item We propose to use the first 100 orbital harmonics in Fourier spectra as features for HBSs to facilitate neural network learning. This approach effectively reduces computational requirements while being suitable for HBSs with arbitrary orbital periods. Furthermore, the method can be extended to other types of periodic variable stars.
	\item RNNs are particularly well suited for processing such time series data. Our results show that an RNN with only two GRU or LSTM layers achieves satisfactory performance, eliminating the need for deeper network architectures. The model achieves 95$\%$ accuracy on the test data set, with an average detection accuracy of 86$\%$ for HBSs from the OGLE, Kepler, and TESS surveys.
	\item Our method can also identify HBSs and eccentric binaries, which are also a class of scientifically valuable eccentric-orbit objects. These two types of objects can be roughly distinguished by the number of eclipses in their phase-folded light curves. For candidate systems that exhibit only a single eclipse, visual inspection is still required to confirm their classification as HBSs. However, the required visual inspection effort has been significantly reduced.
	\item TEOs are a characteristic feature of HBSs. The results show that our method remains effective for HBSs with TEOs.
	\item Four new HBSs, including KICs 4940438, 6794131, 7601633, and 9243795, within the eclipsing binary catalog of Kirk et al., are identified with our method.
\end{enumerate}

%%\section{Code availability}
The neural networks are built using the TensorFlow deep learning API. The trained RNNs and our source code are available via the GitHub repository \footnote{\url{https://github.com/MinyuLi/HBSsNN}}. Our source code includes the RNN definition and the orbital frequency calculation program. Our next work for applying the method to identify new HBSs in TESS data is in preparation.

%%\section{Software and third party data repository citations} \label{sec:cite}
%% Please use the acknowledgment and contribution environments. This will 
%% be anonomyized when the "anonymous" style option is used. 
\begin{acknowledgments}
This work is supported by the Yunnan Fundamental Research Projects (grant Nos. 202501AS070055, 202401AS070046, 202503AP140013, 202301AT070352), the International Partnership Program of Chinese Academy of Sciences (grant No. 020GJHZ2023030GC), the China Manned Space Program with grant No. CMS-CSST-2025-A16, the CAS ``Light of West China" Program, the Basic Research Project of Yunnan Province (grant No. 202201AT070092), the Yunnan Revitalization Talent Support Program, the China Postdoctoral Science Foundation under Grant Number 2025M773194, and the Postdoctoral Fellowship Program of CPSF under Grant Number GZC20252095. The NASA Explorer Program provides funding for the Kepler and TESS missions. We thank the Kepler and TESS teams for their support and hard work. We would like to thank the anonymous referees for many inspiring suggestions that improved this manuscript.
\end{acknowledgments}

\software{ELLC \citep{2016A&A...591A.111M},  
          FNPEAKS (Z. Koo{\l}aczkowski, W. Hebisch, G. Kopacki),
          Tensorflow (\href{https://www.tensorflow.org/}{https://www.tensorflow.org})
          }

%% Appendix material should be preceded with a single \appendix command.
%% There should be a \section command for each appendix. Mark appendix
%% subsections with the same markup you use in the main body of the paper.
%%
%% Each Appendix (indicated with \section) will be lettered A, B, C, etc.
%% The equation counter will reset when it encounters the \appendix
%% command and will number appendix equations (A1), (A2), etc. The
%% Figure and Table counter will not reset.

%%\appendix

%% For this sample we use BibTeX plus aasjournals.bst to generate the
%% the bibliography. The sample7.bib file was populated from ADS. To
%% get the citations to show in the compiled file do the following:
%%
%% pdflatex sample7.tex
%% bibtext sample7
%% pdflatex sample7.tex
%% pdflatex sample7.tex

\bibliography{sample7}{}
\bibliographystyle{aasjournal}

%% This command is needed to show the entire author+affiliation list when
%% the collaboration and author truncation commands are used.  It has to
%% go at the end of the manuscript.
%\allauthors

%% Include this line if you are using the \added, \replaced, \deleted
%% commands to see a summary list of all changes at the end of the article.
%\listofchanges

\end{document}